
%

\input harvmac.tex      

%
%
%
\def\bra#1{{\langle #1 |  }}

\def\bar{\overline}
\def\hat{\widehat}
\def\*{\star}
\def\[{\left[}
\def\]{\right]}
\def\({\left(}		
\def\){\right)}

%
%

\def\frac#1#2{{#1 \over #2}}
\def\inv#1{{1 \over #1}}

\def\d{\partial}

\def\vev#1{\langle #1 \rangle}
\def\ket#1{ | #1 \rangle}
\def\bra#1{\langle #1 |}

\def\2pi{\hbox{$2\pi i$}}

\def\dsl{\raise.15ex\hbox{/}\kern-.57em\partial}
\def\Dsl{\,\raise.15ex\hbox{/}\mkern-.13.5mu D}
%
%

\def\al{\alpha}
\def\ep{\epsilon}

\def\om{\omega}		
\def\sig{\sigma}	

%
%
\def\CA{{\cal A}}		
	\def\CE{{\cal E}}

\def\2pi{\hbox{$2\pi i$}}

\def\dsl{\raise.15ex\hbox{/}\kern-.57em\partial}
\def\Dsl{\,\raise.15ex\hbox{/}\mkern-.13.5mu D}
%
%
%
\font\numbers=cmss12
\font\upright=cmu10 scaled\magstep1
\def\stroke{\vrule height8pt width0.4pt depth-0.1pt}
\def\topfleck{\vrule height8pt width0.5pt depth-5.9pt}
\def\botfleck{\vrule height2pt width0.5pt depth0.1pt}
\def\Zmath{\vcenter{\hbox{\numbers\rlap{\rlap{Z}\kern
0.8pt\topfleck}\kern
2.2pt
                   \rlap Z\kern 6pt\botfleck\kern 1pt}}}
\def\Qmath{\vcenter{\hbox{\upright\rlap{\rlap{Q}\kern
                   3.8pt\stroke}\phantom{Q}}}}
\def\Nmath{\vcenter{\hbox{\upright\rlap{I}\kern 1.7pt N}}}
\def\Cmath{\vcenter{\hbox{\upright\rlap{\rlap{C}\kern
                   3.8pt\stroke}\phantom{C}}}}
\def\Rmath{\vcenter{\hbox{\upright\rlap{I}\kern 1.7pt R}}}
\def\Z{\ifmmode\Zmath\else$\Zmath$\fi}
\def\Q{\ifmmode\Qmath\else$\Qmath$\fi}
\def\N{\ifmmode\Nmath\else$\Nmath$\fi}
\def\C{\ifmmode\Cmath\else$\Cmath$\fi}
\def\R{\ifmmode\Rmath\else$\Rmath$\fi}

\Title{CLNS 94/1302}
{\vbox{\centerline{Quantum Solitons in Non-Linear Optics:  }
\centerline{Resonant Dielectric Media} }}

\bigskip
\bigskip

\centerline{Andr\'e LeClair}
\medskip\centerline{Newman Laboratory}
\centerline{Cornell University}
\centerline{Ithaca, NY  14853}
\bigskip\bigskip

\vskip .3in

It is known that classical electromagnetic radiation
at a frequency in resonance with energy splittings of
atoms in a dielectric medium can be described using the
classical sine-Gordon theory.   In this paper we quantize
the electromagnetic field and compute quantum corrections
to the classical results by using known results from the
sine-Gordon quantum field theory.

\Date{10/94}
%
%
%
%
%
%
\noblackbox

%
%
%
%
%
%
%
%
%
%

\newsec{Introduction}

The importance of integrable non-linear partial differential
equations in classical non-linear optics has been recognized for
some time.  The most well-known example concerns weakly non-linear
dielectric media, where only the first non-linear susceptibility
is considered important. In this situation, the envelope of the electric
field satisfies the non-linear Schrodinger equation.  This was
first understood theoretically by Hasegawa and Tappert\ref\rhas{A. Hasegawa
and F. Tappert, Appl. Phys. Lett. 23 (1973) 142.}.
The solitons predicted in \rhas\ were observed experimentally
in \ref\rmol{L. F. Mollenauer, R. H. Stollen and J. P. Gordon,
Phys. Rev. Lett. 45 (1980) 1095.}.
The occurence of these classical solitons in common optical fibers
promises to revolutionize high-speed telecommunications.

Our interest in this subject concerns the possibly interesting
quantum effects which arise when the electromagnetic field is
quantized.  The latter quantization amounts to studying
an interacting quantum field theory in the classical variables
which satisfy the non-linear differential equation.  These
quantum integrable models have been studied extensively,
and many of their properties have already been exactly computed.
A priori, one expects quantum effects to be small for macroscopically
large objects such as solitons\foot{In fiber optic systems the
soliton consists of a cluster of $10^8$ or more photons.}.
Nevertheless, using known exact results from the quantum non-linear
Schrodinger theory, quantum effects have been predicted and
measured\ref\rnla{S. J. Carter, P. D. Drummund, M. D. Reid and
and P. D. Drummond, Phys. Rev. Lett. 58 (1987) 1841.}\ref\rnlb{P. D.
Drummond and S. J. Carter, J. Opt. Soc. Am. B4 (1987) 1565.}\ref\rnlc{Y.
Lai
and H. A. Haus, Phys. Rev. A40 (1989) 844; H. A. Haus and Y. Lai, J. Opt.
Soc. Am. B7 (1990) 386.}\ref\rnld{M. Rosenbluh and R. M. Shelby,
Phys. Rev. Lett. 66 (1991) 153.}.

It is well-known that
the non-linear dielectric susceptibilities are enhanced when the radiation
is in resonance with the energy splitting of quantum states of the
atoms of the sample.
Near resonance, one is no longer in the weakly non-linear regime
(higher susceptibilities involving higher powers of the electric
field become as important) and the physics is no longer well described
by the non-linear Schrodinger equation.  Remarkably, as was understood
and demonstrated experimentally by McCall and Hahn\ref\trans{S. McCall
and E. L. Hahn, Phys. Rev. 183 (1969) 457.}, the system is well
described classically by another famous integrable equation,
the sine-Gordon (SG) equation.

In this paper, we study the quantum effects which arise when
electric fields in resonance with a dielectric medium are quantized
by using known exact results for the SG quantum field theory.
Though these quantum effects are theoretically interesting,
as we will show, they are unfortunately probably too small to
be measurable at the present time.

\newsec{Classical Theory}

\def\sig{\sigma}
\def\cb{\bar{c}}
\def\bcl{\beta_{cl}}
\def\ce{\CE}
\def\om{\omega}
\def\svev#1{\vev{\sig_{#1}}}

In this section we review the manner in which the classical
sine-Gordon equation arises in resonant dielectric media\trans.

We consider electromagnetic radiation of frequency $\omega$
propagating through a collection of atoms, where the frequency
$\om$ is in resonance with an energy splitting of the atomic
states.  For simplicity, we suppose each atom is a two state
system described by the hamiltonian $H_0$ with
the following eigenstates: $H_0 \ket{\psi_1} = -\inv{2} \hbar
\om_0 \ket{\psi_1}$,
 $H_0 \ket{\psi_2} = \inv{2} \hbar
\om_0 \ket{\psi_2}$, such that $\hbar \om_0$ is the energy difference
of the two states.  In the presence of radiation, the atomic
hamiltonian is
\eqn\eIIi{H_{atom} = H_0 - \vec{p} \cdot \vec{E} , }
where $\vec{p} = e \sum_i \vec{r}_i $ is the electric dipole
moment operator.

We assume the radiation is propagating in the $\hat{z}$ direction,
and $\vec{E} = \hat{n} E(z,t)$, where $ \hat{n} \cdot \hat{z} = 0$.
The only non-zero matrix elements of the operator $\vec{p}\cdot \vec{E}$
can be parameterized as follows:
\eqn\eIIib{
\bra{\psi_2} \vec{p} \cdot \vec{E} \ket{\psi_1}  = p E(z,t) e^{-i\al} ,}
where
$p$ and $\al$ are constants which depend on the atom in question.
(We have assumed spherical symmetry.)  It is convenient to introduce
the Pauli matrices $\sig_i$, and write the hamiltonian as
\eqn\eIIii{
H_{atom} = - \inv{2} \hbar \om_0 \, \sig_3 - E(z,t) (p_1 \sig_1 + p_2
\sig_2 ) , }
where
$p_1 = p \cos \al , ~ p_2 = p\sin \al $.

The dynamics of the system is determined by Maxwell's equations,
\eqn\eIIiii{
\( \d_z^2 - \inv{\cb^2} \d_t^2 \) E(z,t) = \frac{4\pi}{c^2} ~
\d_t^2 P(z,t) ,}
where
$\cb^2 = c^2 / \ep_0$, $\ep_0$ is the ambient dielectric constant,
and $\vec{P} = \hat{n} P(z,t)$ is the dipole moment per unit volume.
The latter polarization can be expressed in terms of the expectations
of the Pauli spin matrices $\vev{\sig_i} = \bra{\psi} \sig_i \ket{\psi}$,
where $\ket{\psi}$ is the atomic wavefunction.  Namely,
\eqn\eIIiv{
\vec{P} = \hat{n}  n \( p_1 \vev{\sig_1} + p_2 \vev{\sig_2} \),}
where
$n$ is the number of atoms per unit volume\foot{To be more precise,
$\vev{\sig_i}$  here represents
average over many atoms in a small  volume, and is thus
a continuous field depending on $z,t$.}.      Thus, in addition
to the Maxwell equation \eIIiii, one has dynamical equations for
the polarization $P(z,t)$ which are determined by Schrodinger's
equation for the atom:
\eqn\eIIv{
i\hbar \,  \d_t \vev{\sig_i} = \bra{\psi} \[ \sig_i , H_{atom} \] \ket{\psi} .}
The latter can be expressed as
\eqn\eIIvi{
\d_t \vev{\sig_i} = \sum_{j,k} \varepsilon_{ijk}  V_j ~ \vev{\sig_k} , }
where
$\varepsilon$ is the completely antisymmetric tensor with
$\varepsilon_{123} = 1$, and
\eqn\eIIvii{
V_1 = \frac{ 2 E(z,t)}{\hbar} p_1 , ~~~~~
V_2 = \frac{ 2 E(z,t)}{\hbar} p_2 , ~~~~~
V_3 = \om_0 .}

To summarize, the dynamics is determined from the coupled equations
of motion \eIIiii\ and \eIIvi, wherein the atoms are treated
quantum mechanically and the radiation is classical.

\def\spar{\svev{\parallel}}
\def\sper{\svev{\perp}}

Let $E(z,t) = \ce (z,t) \cos (\om t - k z)$ where
$\om / k = \cb$ and $\ce (z,t)$ is the envelope of the electric
field.  We will assume the envelope is slowly varying in comparison
to the harmonic oscillations: $\d_t \ce << \om \ce, ~
\d_z \ce << k \ce$.  In this approximation, one finds
\eqn\eIIviii{
\( \d_z^2 - \inv{\cb^2} \d_t^2 \) E(z,t) \approx
\frac{ 2\om}{\cb}
\[ \(\d_z + \inv{\cb} \d_t \) \ce (z,t) \] \sin (\om t - kz ) .}

\def\ar{( \om t - k z + \al) }
Let us define $\spar, \sper$ as follows
\eqn\eIIix{\eqalign{
\spar &= \svev{1} \cos \ar + \svev{2} \sin \ar \cr
\sper &= - \svev{1} \sin \ar + \svev{2} \cos \ar . \cr
}}
We make the further approximation that
$\cos 2\ar  $ terms in the equations of motion
for $\d_t \svev{i}$ can be dropped in comparison to
$\cos \ar$ (and similarly for the sine terms)\foot{It can
be shown in perturbation theory that this is a good approximation
at or near resonance.}.  (This amounts to replacing
$\sin^2 \ar , \cos^2 \ar $ by $1/2$).  One then finds
\eqna\eIIx
$$\eqalignno{
\d_t \spar &= (\om - \om_0 ) \,  \sper  &\eIIx {a} \cr
\d_t \sper &= - \frac{\ce (z,t)}{\hbar} p \, \svev{3}
+ (\om_0 - \om ) \spar  &\eIIx {b} \cr
\d_t \svev{3}  &=  \frac{\ce (z,t)}{\hbar} p\,  \sper  . &\eIIx {c} \cr
}$$
Finally, equations \eIIiii\ and \eIIviii, upon making an approximation
analagous to the slowly varying envelope on the RHS of \eIIiii,
lead to
\eqn\eIIxi{
\[ (\d_z + \inv{\cb} \d_t ) \ce (z,t) \]
\sin (\om t - k z )
= \frac{2\pi }{c^2} \cb n \om p
\( \sin (\om t - kz) \sper - \cos (\om t - kz ) \spar \) . }

In order to solve these equations, note that
$\d_t (\sum_i \svev{i} \svev{i} ) = 0$.  Thus, if the atoms start
out in their ground state with $\svev{3} = 1/2$, one has
$\sum_i \svev{i}^2 = 1/4$ for all time.  On resonance, when
$\om = \om_0$, eq. \eIIx{a} implies $\spar = 0$ for all time.
The constraint $\sum_i \svev{i}^2  = 1/4$
can be imposed with the following parameterization:
\eqn\eIIxii{ \sper = \inv{2} \sin (\bcl \phi (z,t) ) , ~~~~~
\svev{3} = \inv{2} \cos (\bcl \phi (z,t) ).}
The parameter $\bcl$ is arbitrary at this stage, but will be fixed
in the next section.
Equations \eIIx{b,c}\   now imply
\eqn\eIIxiii{
\d_t \phi = - \frac{ p}{\bcl \hbar} ~ \ce (z,t) .}
Inserting \eIIxiii\ into \eIIxi\ and defining $x=2z - \cb t$,
one obtains the SG equation:
\eqn\eIIxiv{
\( \d_t^2 - \cb^2 \d_x^2 \) \phi = - \frac{\mu^2}{\bcl}
\sin (\bcl \phi ) , }
where
\eqn\eIIxv{
\mu^2 = \frac{2\pi n p^2 \om}{\hbar \ep_0} . }

\newsec{Quantum Effects}

In this section we proceed to quantize the electromagnetic field.
In order to do this, the SG field $\phi$ must be properly normalized
such that the energy of soliton solutions corresponds to the
physical energy;  this amounts to properly fixing the constant
$\bcl$.

The action which gives the classical SG equation of motion is
\eqn\eIIIi{
S_{SG} = \inv{\cb} \int dx dt \(
\inv{2} (\d_t \phi)^2 - \frac{\cb^2}{2} (\d_x \phi )^2
+ \frac{\mu^2}{\bcl^2} ~ \cos (\bcl \phi ) \) . }
On the other hand, the properly normalized Maxwell action is
\eqn\eIIIii{
S_{\rm Maxwell} = \inv{c^2} \int d^3 x dt ~ \( - \inv{4}
F_{\mu\nu} F^{\mu\nu} + ... \)  ~~~=
\inv{c^2} \int d^3 x dt ~ \inv{2} (\d_t A )^2 + .... }
where the vector potential is $\vec{A} = \hat{n} A $ and
as usual $\vec{E} = -\inv{c} \d_t \vec{A} $.  (One can show
that $A_0$ can be set to zero.)  To normalize the field
$\phi$, one needs only compare the kinetic terms in
\eIIIi\ and \eIIIii.  The dimensional reduction is made by
assuming simply that $A$ is independent of $y,z$ and
$\int dydz = \CA$, where $\CA$ is an effective cross-sectional
area perpendicular to the direction of propagation (as in the
cross-sectional area of a fiber).  From \eIIxiii\ one has
\eqn\eIIIiii{
S_{\rm Maxwell} = \frac{\CA}{2} \( \frac{\bcl \hbar}{p} \)^2
\int dx dt ~  \inv{2} (\d_t \phi )^2 + .... }
Comparing with \eIIIi, one fixes $\bcl^2 = 2p^2 /(\CA \hbar^2 \cb)$.

Finally, as is conventionally done, we rescale $\phi \to
\sqrt{\hbar} \phi$, so that $S_{SG} / \hbar$ takes the form \eIIIi
with $\bcl$ replaced by $\beta$, where
\eqn\eIIIvi{
\beta^2 = \hbar \bcl^2 = \frac{2 p^2 \sqrt{\ep_0}}{\CA \hbar c} . }
The constant $\beta$ is the conventional dimensionless coupling
constant in the quantum SG theory, and is allowed to be in
the range $0\leq \beta^2 \leq 8\pi$, where
$\beta^2 = 8\pi$ corresponds to a  phase
transition\ref\rcole{S. Coleman, Phys. Rev D11 (1975) 2088.}.
The limit where the radiation
is classical but the atom is still quantum mechanical corresponds
to the limit $\beta \to 0$.

\def\bep{\frac{\beta^2}{8\pi}}

The magnitude of the quantum corrections to classical results is
determined by the parameter $\beta^2/8\pi$.  One can obtain an
order of magnitude estimate by taking $p \approx e R_{\rm bohr}$,
where $R_{\rm bohr}$ is the Bohr radius.  One finds
\eqn\eIIIvii{
\bep \approx  10^{-21} \frac{\sqrt{\ep_0}}{\CA}, }
where $\CA$ is in ${\rm cm}^2$.  Disappointingly, for
realistic $\CA$, the constant $\beta^2 /8\pi $
is exceedingly small. For the remainder
of this paper we describe quantum effects that are at least
in principle measureable, though in practice probably too
small to observe.

The classical soliton solutions to the SG equation are characterized
by a topological charge $T= \pm 1$, where
$$T= \frac{\beta}{2\pi }  (\phi (x= \infty) - \phi (x= -\infty ) ) .$$
Solitons
of either charge correspond to solutions where at fixed $z$ the atoms in
the far past are in their ground state, and are all in their excited
state at some intermediate time:  $\svev{3}_{t= \pm \infty} = 1/2$.
These classical solitons have been observed experimentally in \trans.
What distinguishes solitons ($T=1$) from antisolitons ($T=-1 $)
is the sign of the envelope of the electric field.  From the known
classical soliton solutions and \eIIxiii\ one finds for
$T= \pm 1$,
\eqn\eIIIviii{
\ce (x,t) = \pm \frac{2 \hbar \mu}{p} \sqrt{ \frac{\cb + v}{\cb -v} }
\(  \cosh \(  \frac{\mu (x-vt)}{\sqrt{\cb^2 - v^2}} \) \)^{-1}  .  }
Thus the electric fields for the soliton versus the antisoliton
are out of phase by $\pi$.

The classical soliton mass is given by $M_s = 8m/ \beta^2$, where
$m = \mu \hbar /\cb^2 $.  From \eIIIvi\ one can express
$\beta^2 /8\pi$ in terms of the classical soliton mass:
\eqn\eIIIix{
\bep = 8 \sqrt{\ep_0} \( \frac {\hbar \om}{M_s c^2} \)
\( \frac{\hbar n \CA c}{M_s c^2}\) . }
Thus,
\eqn\eIIIx{
\bep \sim ~~  \inv{N_\gamma} \( \frac{\lambda_c}{L_{\rm atom}} \),
}
where $N_\gamma = \hbar \om / M_s c^2$ roughly corresponds classically
to the number of photons that comprise the soliton,
$\lambda_c$ is the Compton wavelength of the soliton, and
$L_{\rm atom} = 1/n\CA$ is the inter-atomic spacing.  The above
equation summarizes where one expects quantum effects to be
important: when the soliton is comprised of small numbers of photons,
or when the Compton wavelength is large compared to the space
between the atoms.

We now derive quantum corrections to the frequency and density
dependence of the soliton mass.  Classically, $M_s \propto
\sqrt{\om n}$.  In the quantum theory, short distance singularities
are removed by suitably normal ordering the $\cos (\beta \phi )$
potential\rcole:
$$\frac{\mu^2}{\beta^2}  \cos (\beta \phi )  \to \lambda
: \cos (\beta \phi ) :$$  The anomalous scaling dimension
of the operator $:\cos (\beta \phi ) :$ is $\beta^2 / 4 \pi $,
so that $\lambda$ has mass dimension $2 (1- \bep )$.
Therefore,
\eqn\ems{M_s = c(\beta ) ( \sqrt{\lambda})^{1/(1-\bep)} }  for
some constant $c(\beta )$.  The latter constant was computed
exactly in \ref\ralzam{Al. B. Zamolodchikov, {\it Mass Scale in
Sine-Gordon and its Reductions}, Montpellier preprint
LPM-93-06, 1993.}.  Since $\lambda \propto
\om n$, one finds
\eqn\eIIIxi{
M_s  \propto
\sqrt{\om n} \( 1 + \frac{\beta^2}{16\pi} \log (\om n )
+ O(\beta^4 ) \). }

The classical scattering matrix for the solitons has been computed
by comparing N-soliton solutions in the far past and far future\ref\rkor{V.
E. Korepin and L. D. Faddeev, Theor. Mat. Fiz 25 (1975) 147.}.
The exact quantum S-matrix is also known\ref\rzz{A. B. Zamolodchikov
and Al. B. Zamolodchikov, Ann. Phys. 120 (1979) 253.}.  Since
$\beta^2 / 8\pi $ is small here,
the quantum corrections to classical scattering
are most easily determined by incorporating the one-loop corrections
to the classical scattering, which amounts to the
replacement $\beta^2 \to \gamma = \beta^2 / (1- \bep )$. The
necessary formulas can be found in the above papers.

The most interesting aspect of the quantum scattering of solitons
is
that the so-called reflection amplitude is a purely quantum effect,
analagous to barrier penetration \ref\rkorb{V. E. Korepin,
Teor. Mat. Fiz. 34 (1978) 3.}.
Namely, one considers an in-state consisting of a soliton
of momentum $p_1$ and an antisoliton of momentum $p_2$ which scatters
into an out state where the momenta $p_1$ and $p_2$ are interchanged.
The $T= +1$
soliton is thus reflected back with momentum equal to that of
the incoming $T=-1$ antisoliton.  In the semi-classical approximation,
this reflection amplitude is
\eqn\eIIIxii{
S_R (\theta ) = \inv{2} \( e^{16\pi^2 i /\gamma } - 1 \)
e^{- \frac{8\pi }{\gamma} |\theta | }  ~ S(\theta )
}
where
\eqn\eIIIxiii{
S(\theta ) = \exp \( \frac{8}{\gamma} \int_0^\pi
d\eta
\log \[ \frac{e^{\theta - i \eta} + 1}{e^\theta + e^{-i\eta} } \]
\), }
and
$\theta = \theta_1 - \theta_2 $, where
$p_{1,2} = M_s \sinh \theta_{1,2} $.
Since $1/\gamma \approx 1/\beta^2$ is very large, the oscillatory
factor $\exp( 16\pi^2 i/\gamma ) - 1$ in \eIIIxii\
will make the detection of reflection processes difficult,
since even in
a small range of $\beta^2$, this factor averages to zero.

\bigskip

\centerline{\bf Acknowledgements}

It is a pleasure to thank Peter Lepage and
Sergei Lukyanov for useful discussions.
This work is supported by an Alfred P. Sloan Foundation fellowship,
and the National Science Foundation in part through the
National Young Investigator program.

\listrefs
\end